\documentclass[10pt, draftclsnofoot, onecolumn]{IEEEtran}
\IEEEoverridecommandlockouts

\usepackage{cite}
\usepackage{amsmath,amssymb,amsfonts}
\usepackage{algorithmic}
\usepackage{graphicx}
\usepackage{textcomp}
\usepackage{psfrag,caption,subcaption}
\usepackage{xcolor}
\def\BibTeX{{\rm B\kern-.05em{\sc i\kern-.025em b}\kern-.08em
    T\kern-.1667em\lower.7ex\hbox{E}\kern-.125emX}}
    
\input{mysymbol.sty}

\title{Channel Estimation for RIS-assisted Millimeter-wave MIMO Systems}

\author{Battu Deepak,  R.S. Prasobh Sankar, and Sundeep Prabhakar Chepuri, \IEEEmembership{Member,~IEEE}

\thanks{The authors are with the Department of ECE, Indian Institute of Science, Bangalore, India. Emails: \{deepakbattu, prasobhr, spchepuri\}@iisc.ac.in.}

}

\begin{document}

\maketitle

\begin{abstract}

Channel estimation in reconfigurable intelligent surface (RIS) assisted multiple input multiple output (MIMO) communication systems is challenging as pilots cannot be decoded at or transmitted from a RIS with only passive elements. We assume an angular model for the line-of-sight channel between the user equipment (UE) and the RIS and between the RIS and the base station (BS) with no direct path between the UE and BS. In this letter, assuming that we can compute the RIS-BS channel (up to a complex path gain) from their known locations, we first discuss the ambiguity involved in estimating the UE-RIS channel. We propose a multiple channel sounding procedure in which we observe the channel through different RIS phase shifts and present an algorithm for channel estimation to resolve this ambiguity. Through simulations, we show that the proposed method is comparable to an oracle estimator, which perfectly knows the BS, UE, and RIS locations.

\end{abstract}

%Although the LoS channel (up to the complex path gain) can be computed when the locations of the BS, RIS, and UE are known, the RIS-UE channel needs to be estimated as the UE location is usually not known.
\begin{IEEEkeywords}
Channel estimation, direction estimation, MIMO, millimeter-wave, reconfigurable intelligent surfaces.
\end{IEEEkeywords}

\section{Introduction} \label{sec:introduction}
The availability of unused frequency resources at millimeter-wave (mmWave) frequencies and the ever-increasing demand for higher data rates make mmWave MIMO systems a natural choice for the next generation wireless communications~\cite{rappaport2013millimeter,ghosh2014millimeter}. One of the major challenges in operating at mmWave frequencies is the heavy path loss with weak or no line-of-sight (LoS) components~\cite{wan2020broadband,mo2014channel}. Therefore ensuring a strong LoS component is necessary to establish a reliable communication link.

Reconfigurable intelligent surfaces (RISs) can be used to control the propagation environment and ensure a reasonable LoS communication link~\cite{basar2019wireless, ozdogan2019intelligent, ozdogan2020using, bjornson2020intelligent, najafi2020physicsbased, arun2019rfocus}. RIS is a two-dimensional structure consisting of many passive sub-wavelength elements, which act as diffuse scatterers. By controlling the surface impedance of these elements, we can steer and focus the energy of the electromagnetic wave impinging on the RIS to any desired direction~\cite{najafi2020physicsbased}. We can view these elements as phase shifters, each of which can be controlled independently and remotely  via a low-rate link~\cite{basar2019wireless}. The channel estimation problem in a RIS-assisted MIMO system amounts to estimating the MIMO channel between the transmitter, e.g., a user equipment~(UE), and the RIS and the MIMO channel between the RIS and the receiver, e.g., a base station~(BS). When the channel is known, the RIS phase shifts can be designed to form beams in any desired direction. However, due to the passive nature of the RIS with no pilot decoding or transmission capabilities, we cannot readily use conventional pilot-based channel estimation techniques.

\subsection{Related prior works} \label{subsec:intro:priorworks}

Recent works on channel estimation in RIS-assisted mmWave MIMO systems can be broadly classified as follows.
\begin{itemize}
 \item {\bf RIS with a few active elements}: Interleaving RISs with a small number of active elements to decode pilots allows a separate estimation of the UE-RIS and the RIS-BS channels.  With such a RIS configuration, channel estimation algorithms for non-parametric and angular channel models were proposed in~\cite{taha2019enabling} and~\cite{xiao2021low_complexity}, respectively. 
 
\item {\bf RIS with passive-only elements}: RIS with passive-only elements are desirable because of their lower power consumption and hardware costs. Existing techniques estimate a cascaded channel, i.e., the overall channel between the UE and BS via the RIS exploiting the inherent sparsity in the angular domain in mmWave MIMO channels~\cite{wang2020compressed,he2020channelirw,he2020channelanm}.  In~\cite{he2020cascaded}, a bilinear matrix factorization technique was proposed to factorize the cascaded channel matrix into low-rank UE-RIS and RIS-BS channel matrices. 

\end{itemize}

Uniquely estimating the UE-RIS and RIS-BS channels is challenging because of the ambiguity involved in resolving the complex path gains and angles (see Sec.~\ref{sec:uniqueness} for details). Unlike the RIS-BS channel, which is mostly time-invariant due to the fixed locations of the RIS and the BS, the UE-RIS channel can be time-varying due to the mobility of the UE~\cite{xiao2021low_complexity}. Separately estimating the UE-RIS and RIS-BS channels allows us to exploit any difference in the time scales of variation and avoid unnecessary estimation of the cascaded channel or its factorization each time. In addition, in a multi-user setting, estimating only the UE-RIS channels with a shared RIS-BS channel is computationally less expensive.  The RIS-BS channel is usually LoS and is completely characterized (up to a complex path gain) by the known locations of the RIS and BS~\cite{wan2020broadband}. Therefore, the channel estimation amounts to estimating the UE-RIS channel, or equivalently estimating the complex path gain and localizing the UE, which are the main goals of this work. 

%Separately estimating the UE-RIS and BS-RIS channels allows us to exploit any difference in the time scales of variation and avoid unnecessary estimation of the cascaded channel or its factorization each time. This is also advantageous in a multi-user scenario since all users share the same BS-RIS channel~\cite{xiao2021low_complexity}.
%Separate estimation is also beneficial in multi-user scenarios where all users share the same BS-RIS channel.

\subsection{Main results and contributions} \label{subsec:intro:mainresults}

In this work, we propose a pilot-based uplink channel estimation algorithm to estimate the unknown channel parameters, namely, the angle of departure (AoD) at the UE, azimuth and elevation angles of arrival (AoA) at the RIS, and the overall complex path gain of the UE-RIS-BS link. 

One of the main results of this work is, we show that at least two channel soundings with different RIS phase shifts are required to estimate the complex path gain and the angles uniquely, and thus the  UE-RIS channel.  The channel estimation is carried out entirely at the BS without any direct link (low-rate feedback or communication) between the BS and  UE. Through simulations, we demonstrate that the proposed technique performs similarly in terms of normalized mean squared error and average spectral efficiency to an oracle estimator, which assumes perfect knowledge of the angles associated with the  UE-RIS and RIS-BS links.  

\begin{table}[t]
\begin{center}
\small
 \begin{tabular}{l  l} 
 \hline
 Symbol & Definition \\ [0.5ex] 
 \hline\hline
 $\theta_{\rm u}$ &AoD at the UE \\ 
 %\hline
 $(\phi_{\rm u}, \psi_{\rm u})$ & Elevation and azimuth AoA at the RIS from the UE  \\
 %\hline
 $\theta_{\rm b}$ & AoA at the BS  \\
 %\hline
 $(\phi_{\rm b}, \psi_{\rm b})$ & Elevation and azimuth AoD from the RIS to the BS  \\
 %\hline
 $g$ & Complex path gain \\
  \hline
\end{tabular}
\end{center}
\vspace{-2mm}
\caption{Channel parameters.}
\label{table:1}
%\vspace*{-4mm}
\vspace*{-7mm}
\end{table}

\section{RIS-assisted MIMO channel}~\label{sec:prob_model}

In this section, we introduce the signal model and describe the main challenge involved in estimating the UE-RIS channel.

\subsection{Angular MIMO channel model} 
Consider a MIMO communication system having a BS with a uniform linear array (ULA) of $N_{\rm{b}}$ antennas, a UE with a ULA of $N_{\rm{u}}$ antennas, and a RIS with a uniform planar array (UPA) of $N_{\rm{r}}$ passive phase shifters. Let us define the array response vector of a ULA  with $N$ antennas  as
$\mathbf{a}(u,N) = \begin{bmatrix} 1 & e^{-j 2\pi d u/\lambda} & \cdots & e^{-j (N-1) 2\pi d u/\lambda} \end{bmatrix}^T.$ Here, $u$ is the direction cosine, $d$ is the inter-element spacing, and $\lambda$ is the signal wavelength. Then the array response vectors of the arrays at the BS, UE, and RIS are, respectively, defined as $\mathbf{a}_{\rm B}(\theta) = \mathbf{a}({\rm sin(\theta)}, N_{\rm b})$, $\mathbf{a}_{\rm U}(\theta) = \mathbf{a}({\rm sin(\theta)}, N_{\rm u})$, and $\mathbf{a}_{\rm R}(\phi,\psi) = \mathbf{a}({\rm sin(\phi) sin(\psi)} , N_{\rm x}) \otimes \mathbf{a}({\rm sin(\phi) cos(\psi)} , N_{\rm y})$. Here, $N_{\rm x}$ and $N_{\rm y}$ denote the number of RIS elements in the horizontal and vertical directions, respectively, so that $N_{\rm r} = N_{\rm x}N_{\rm y}$, and $\otimes$ denotes the Kronecker product.

We assume that the direct path between the BS and UE is blocked and that the RIS is used to establish a non-direct UE-BS link. Let us denote the UE-RIS and RIS-BS MIMO channel matrices  by $\mathbf{H}_{\rm{ur}} \in \mathbb{C}^{N_{\rm r} \times N_{\rm u}}$ and $\mathbf{H}_{\rm{rb}} \in \mathbb{C}^{N_{\rm b} \times N_{\rm r}}$, respectively. Let us collect the phase shifts in the diagonal matrix $\boldsymbol{\Omega} \in \mathbb{C}^{N_{\rm r} \times N_{\rm r}}$. Usually, the BS and RIS are situated in environments with limited local scattering, for which $\mathbf{H}_{\rm rb }$ will be an LoS channel matrix~\cite{wan2020broadband}. We also assume that the UE-RIS link is LoS. Although this is a simplifying assumption, it is reasonable because of the excessive path loss at mmWave frequencies~\cite{wan2020broadband,mo2014channel}.  Under these assumptions, both $\mathbf{H}_{\rm rb}$ and $\mathbf{H}_{\rm ur}$ are rank-1 matrices defined as
\[
    \mathbf{H}_{\rm{ur}} = g_{\rm{ur}}\mathbf{a}_{\rm R}(\phi_{\rm u},\psi_{\rm u})\mathbf{a}_{\rm U}^{H}(\theta_{\rm u}), 
    \,\,\,      \mathbf{H}_{\rm{rb}} = g_{\rm{rb}}\mathbf{a}_{\rm B}(\theta_{\rm b})\mathbf{a}_{\rm R}^{H}(\phi_{\rm b},\psi_{\rm b}),
\]
where $g_{\rm{ur}}$ and $g_{\rm{rb}}$ denote the complex path gains, $\theta_{\rm u}$ is the AoD at the UE, and  $\phi_{\rm u}$ and $\psi_{\rm u}$ are, respectively, the elevation and azimuth angles at the RIS made by the LoS path arriving from the UE, $\theta_{\rm b}$ is the AoA at the BS, and $\phi_{\rm b}$ and $\psi_{\rm b}$ are the angles made by the LoS path departing from the RIS towards the  BS in the elevation and azimuth directions, respectively. Then the cascaded MIMO channel matrix is given by
\begin{equation} \label{eq:model_1}
    \mathbf{H} = g \mathbf{a}_{\rm B}(\theta_{\rm b})\mathbf{a}_{\rm R}^{H}(\phi_{\rm b},\psi_{\rm b}) \boldsymbol{\Omega} \mathbf{a}_{\rm R}(\phi_{\rm u},\psi_{\rm u})\mathbf{a}_{\rm U}^{H}(\theta_{\rm u}),
\end{equation}
where $g = g_{\rm{rb}}g_{\rm{ur}}$ is the overall complex path gain.  
%Here, $\mathbf{a}_{\rm U}(\theta) \in \mathbb{C}^{N_{\rm u}}$, $\mathbf{a}_{\rm B}(\theta) \in \mathbb{C}^{N_{\rm b}}$, $\mathbf{a}_{\rm R}(\phi, \psi) \in \mathbb{C}^{N_{\rm r}}$ are the array response vectors at the UE, BS and RIS, respectively.
The channel parameters, which we frequently refer, are summarized in~Table~\ref{table:1}.

The main aim of this work is to estimate the parameters $\{\theta_{\rm u}, \phi_{\rm u}, \psi_{\rm u}, g\}$ assuming that the locations of the BS and RIS are known. In other words, we localize the UE by finding the directions $\theta_{\rm u}$ and $(\phi_{\rm u},\psi_{\rm u})$.

\subsection{Uniqueness} \label{sec:uniqueness}

Let us decompose the channel matrix as ${\mathbf{H}} =  c \mathbf{a}_{\rm B}(\theta_{\rm b}) \mathbf{a}_{\rm U}^{H}(\theta_{\rm u})$
with $c = g\mathbf{a}_{\rm R}^{H}(\phi_{\rm b},\psi_{\rm b}) \boldsymbol{\Omega} \mathbf{a}_{\rm R}(\phi_{\rm u},\psi_{\rm u})$. 
Since the scalar parameter $c$ is a product of two complex numbers $\mathbf{a}_{\rm R}^{H}(\phi_{\rm b},\psi_{\rm b}) \boldsymbol{\Omega} \mathbf{a}_{\rm R}(\phi_{\rm u},\psi_{\rm u})$ and $g$, it is not possible to uniquely identify $(\phi_{\rm u},\psi_{\rm u})$ and $g$ from ${\bf H}$ as there are different set of angles $(\phi_{\rm u},\psi_{\rm u})$ and $g$ that result in the same product $c$.  This ambiguity makes the UE-RIS channel estimation challenging.

To resolve this ambiguity, suppose we sound the channel with two different phase shifts $\boldsymbol{\Omega}_{1}$ and $\boldsymbol{\Omega}_{2}$ to obtain
\begin{align} 
    c_{i} {=}  \,\, g \mathbf{a}_{\rm R}^{H}(\phi_{\rm b},\psi_{\rm b}) \boldsymbol{\Omega}_{i} \mathbf{a}_{\rm R}(\phi_{\rm u},\psi_{\rm u}),  \quad\label{eq:ci}
\end{align}
for $ i=1,2$. By computing $c_{1}/c_{2}$, we can eliminate $g$ in \eqref{eq:ci} as
\begin{equation} \label{eq:modified_stage2}
    \mathbf{a}_{\rm R}^{H}(\phi_{\rm b},\psi_{\rm b})( c_{1}\boldsymbol{\Omega}_{2} - c_{2}\boldsymbol{\Omega}_{1})\mathbf{a}_{\rm R}(\phi_{\rm u},\psi_{\rm u}) = 0.
\end{equation}
Therefore, it is immediately clear that {\it at least two} channel soundings are required to resolve the ambiguity. For robustness against noise, we  extend~\eqref{eq:modified_stage2} to {\it multiple channel soundings} with different RIS phase shift matrices and uplink pilots as discussed next.

\subsection{Uplink training} \label{sec:signal_model}

Let $\mathbf{S} \in \mathbb{C}^{N_{\rm u} \times M}$ be the uplink pilot matrix transmitted from the UE with $M$ denoting the number of channel uses. We perform multiple channel soundings with the same $\mathbf{S}$ over $L$ training blocks, but with different phase shifts $\boldsymbol{\Omega}_{i}$  for $i=1,2,\ldots,L$. 
The signal received at the BS during the $i$th training block, $\mathbf{X}_i \in \mathbb{C}^{N_{\rm b}\times M}$, is given by
\begin{equation} \label{symb1}
    \mathbf{X}_i = \mathbf{H}_i \mathbf{S} + \mathbf{N}_i, \quad i=1,\ldots, L,
\end{equation}
where $\mathbf{H}_{i} = g \mathbf{a}_{\rm B}(\theta_{\rm b})\mathbf{a}_{\rm R}^{H}(\phi_{\rm b},\psi_{\rm b}) \boldsymbol{\Omega}_{i} \mathbf{a}_{\rm R}(\phi_{\rm u},\psi_{\rm u})\mathbf{a}_{\rm U}^{H}(\theta_{\rm u})$, $\mathbf{N}_i \in \mathbb{C}^{N_{\rm b}\times M} $ is the noise matrix whose each entry follows a complex Gaussian distribution as $[\mathbf{N}_i]_{mn} \sim \mathcal{C}\mathcal{N}(0,1)$. 
Without loss of generality, we consider the pilot matrix to be orthogonal with $M = N_{\rm u}$ so that $\mathbf{S}\mathbf{S}^{H} = \mathbf{S}^{H}\mathbf{S} = PN_{\rm u}^{-1}\mathbf{I}_{N_{\rm u}}$. Here, $P$ is the signal-to-noise ratio (SNR).

\section{Proposed algorithm} \label{sec:ch_estm_algo}

In this section, we develop an algorithm to estimate the UE-RIS channel based on the observations from multiple channel soundings. The algorithm comprises of three stages to estimate the {\it AoD at the UE}, {\it AoA from UE at RIS}, and  the {\it path gain}. Before describing the algorithm, let us first obtain a coarse estimate of the channel in each block as
\begin{equation} \label{eq:hnoisegen}
    \Tilde{\mathbf{H}}_{i} = \frac{N_{\rm u}}{P}\mathbf{X}_{i}\mathbf{S}^{H} = \mathbf{H}_{i} + \mathbf{W}_{i}, \quad i=1,2,\ldots,L,
\end{equation}
where $\mathbf{W}_{i} = N_{\rm u} P^{-1}\mathbf{N}_{i}\mathbf{S}^{H}$ is the noise term after pilot removal. 

%Now, to estimate the channel matrix, we  estimate $\theta_{\rm u}$, $\{\phi_{\rm u},\psi_{\rm u}\}$, and $g$ using a three-step process as described next.

\subsection{Estimation of the AoD at the UE}

To estimate the AoD at the UE, i.e., $\theta_{\rm u}$, we use MUSIC~\cite{vantrees2002optimum}, which is a subspace-based direction finding method. In the absence of noise, we have $\mathcal{R}(\Tilde{\mathbf{H}}_{i}^{H}) = \mathcal{R}(\mathbf{a}_{\rm U}(\theta_{\rm u})) \> \forall \> i=1,2,\ldots,L$. Here, $\mathcal{R}({\bf H})$ denotes the column span of ${\bf H}$.  We can estimate $\mathcal{R}(\mathbf{a}_{\rm U}(\theta_{\rm u}))$ from the observations from all the $L$ blocks by forming the rank-1 matrix ${\mathbf{G}} \in \mathbb{C}^{N_{\rm u} \times LN_{\rm b}}$ as
\begin{equation}
    \mathbf{G} = \begin{bmatrix} \Tilde{\mathbf{H}}_{1}^{H} & \Tilde{\mathbf{H}}_{2}^{H} & \ldots & \Tilde{\mathbf{H}}_{L}^{H}     \end{bmatrix},
\end{equation}
where $\mathcal{R}(\mathbf{G}) = \mathcal{R}(\mathbf{a}_{\rm U}(\theta_{\rm u}))$. Let us denote the subspace orthogonal to the one-dimensional subspace $\mathcal{R}(\mathbf{a}_{\rm U}(\theta_{\rm u}))$ 
by $\mathbf{U}$ so that  $\mathbf{U}^H\mathbf{a}_{\rm U}(\theta_{\rm u}) = {\bf 0}$. We may compute $\mathbf{U}$ using the singular value decomposition (SVD) of $\mathbf{G}$ and by considering the left singular vectors corresponding to all but its largest singular value. In the presence of noise, we minimize $\vert \vert \mathbf{U}
^{H}\mathbf{a}_{\rm U}(\theta) \vert \vert^{2}$ with respect to $\theta$. The estimate of the AoD at the UE, denoted by $\hat{\theta}_{\rm u}$, can be obtained from the peak of the pseudo spectrum 
\begin{equation} \label{eq:theta_est}
    \mathcal{P}_{\rm UE}(\theta) = {\vert \vert \mathbf{U}^{H}\mathbf{a}_{\rm U}(\theta) \vert \vert^{-2}}
\end{equation}
by sweeping over $\theta$. 

Although $\theta_{\rm u}$ may also be estimated using any other standard direction finding technique (e.g., sparse recovery or maximum likelihood estimators), the estimation of $(\phi_{\rm u},\psi_{\rm u})$ that we discuss next is more challenging as we do not have access to the measurements at the RIS. 
%Next, we present the algorithm to separately estimate the complex path gain and the AoAs at the RIS using $\hat{\theta}_{\rm u}$.

\subsection{Estimation of the AoA from the UE at the RIS}

Given $\hat{\theta}_{\rm u}$, we can obtain noisy measurements of $c_{i}$ by multiplying $\tilde{\mathbf{H}}_{i}$ in \eqref{eq:hnoisegen} from the left with $\mathbf{a}_{\rm B}^H(\theta_{\rm b})$ and from the right with $\mathbf{a}_{\rm U}(\hat{\theta}_{\rm u})$ as $\mathbf{a}_{\rm B}^H(\theta_{\rm b})\tilde{\mathbf{H}}_{i}\mathbf{a}_{\rm U}(\hat{\theta}_{\rm u})$ to obtain [cf. \eqref{eq:ci}]
\begin{equation}
     \mathbf{a}_{\rm R}^{H}(\phi_{\rm b},\psi_{\rm b})( c_{i}\boldsymbol{\Omega}_{j} - c_{j}\boldsymbol{\Omega}_{i})\mathbf{a}_{\rm R}(\phi_{\rm u},\psi_{\rm u}) = e_i, 
     \label{eq:noisyci}
\end{equation}
for $i,j=1,2, \ldots, L$. Here, $e_i$ is the error due to the noise~$\mathbf{W}_i$. 

Using \eqref{eq:noisyci}, we pose the problem of estimating $(\phi_{\rm u}, \psi_{\rm u})$ as the problem of computing the solution to a system of non-linear equations. Although there are multiple ways to form the required system of equations, we may stack \eqref{eq:noisyci} for $i=1$ and $j=2,3,\ldots, L$ to form 
\begin{equation} \label{phipsicombined}
\underbrace{  \begin{bmatrix} \mathbf{a}_{\rm R}^{H}(\phi_{\rm b},\psi_{\rm b})( c_{1}\boldsymbol{\Omega}_{2} - c_{2}\boldsymbol{\Omega}_{1}) \\
\mathbf{a}_{\rm R}^{H}(\phi_{\rm b},\psi_{\rm b})( c_{1}\boldsymbol{\Omega}_{3} - c_{3}\boldsymbol{\Omega}_{1}) \\
\vdots \\
\mathbf{a}_{\rm R}^{H}(\phi_{\rm b},\psi_{\rm b})( c_{1}\boldsymbol{\Omega}_{L} - c_{L}\boldsymbol{\Omega}_{1})\end{bmatrix} }_\text{$\Tilde{\mathbf{A}} \in \mathbb{C}^{(L-1)\times N_{\rm r}}$} \mathbf{a}_{\rm R}(\phi_{\rm u},\psi_{\rm u})  = {\bf e},\end{equation}
where ${\bf e}$ denotes the error vector. Since we do not know the direction of the UE during the channel estimation phase, $\boldsymbol{\Omega}_{i}$, $i=1,2,\ldots,L$ are randomly chosen such that all the directions are excited. Thus the matrix $\Tilde{\mathbf{A}}$ will be full row rank.  

Let us denote the basis for $\mathcal{R}(\Tilde{\mathbf{A}}^{H})$ as $\mathbf{Q} \in \mathbb{C}^{N_{\rm r}\times (L-1)}$, which may be obtained using the SVD of $\Tilde{\mathbf{A}}^{H}$ and by considering the left singular vectors corresponding to the $(L-1)$ largest singular values of $\Tilde{\mathbf{A}}^{H}$.  In the noiseless setting with ${\bf e} = {\bf 0}$ (as in \eqref{eq:ci}), $\mathbf{a}_{\rm R}(\phi_{\rm u},\psi_{\rm u})$ lies in the null space of $\Tilde{\bf A}$ so that $\vert\vert \mathbf{Q}^{H}\mathbf{a}_{\rm R}(\phi_{\rm u},\psi_{\rm  u})  \vert\vert^{2} = 0$. Therefore, in presence of noise, minimizing $\vert\vert \mathbf{U}_{s}^{H}\mathbf{a}_{\rm R}(\phi,\psi) \vert\vert^{2}$ with respect to $(\phi,\psi)$ amounts to solving the system of non-linear equations in \eqref{phipsicombined} in the least squares sense. Thus, we can obtain the estimates of the AoA at the RIS, $(\hat{\phi}_{\rm u},\hat{\psi}_{\rm u})$, by computing the locations of the peaks of the pseudo spectrum 
\begin{equation} \label{eq:phi_psi_est}
    \mathcal{P}(\phi,\psi) = {\vert\vert \mathbf{U}_{s}^{H}\mathbf{a}_{\rm R}(\phi,\psi)  \vert\vert^{-2}},
\end{equation}
by sweeping over $(\phi,\psi)$

% The angular resolution of the estimators from \eqref{eq:theta_est} and \eqref{eq:phi_psi_est} depends on dimension on search grid.

% Next, we estimate the complex path gain using the estimates $\hat{\theta}_{\rm u}$ and $\{\hat{\phi}_{\rm u}, \hat{\psi}_{\rm u}\}$.

%\vspace{-0.05in}

\subsection{Estimation of the path gain} \label{sec:estm_g}

Once we have the estimates of the AoA and AoD available, estimating the path gain $g$ can be done using least squares. Let us define the following $N_{\rm b}$-length vectors
$\mathbf{b}_{i} = \Tilde{\mathbf{H}}_{i}\mathbf{a}_{{\rm U}}({\theta}_{\rm u})$ and 
${\bf v}_i = N_{\rm u}\mathbf{a}_{\rm B}(\theta_{\rm b})\mathbf{a}_{\rm R}^{H}(\phi_{\rm b},\psi_{\rm b}) \boldsymbol{\Omega}_{i} \mathbf{a}_{\rm R}(\phi_{\rm u},\psi_{\rm u})$. Then, from \eqref{eq:model_1} and \eqref{eq:hnoisegen}, we have
\begin{align} \label{bdef1}
 \mathbf{b}_{i}  &= g{\bf v}_i + \mathbf{z}_{i},  \> i=1,2,\ldots,L,       
\end{align}
where $\mathbf{z}_{i} = \mathbf{W}_{i}\mathbf{a}_{{\rm U}}({\theta}_{\rm u})$ is the noise vector.
%
%\begin{equation}
%    \mathbf{b}_{i} \overset{\Delta}{=} \Tilde{\mathbf{H}}_{i}\mathbf{a}_{{\rm u}}({\theta}_{\rm u}), \> i=1,2,...,L.
%\end{equation}
%If we assume that $\hat{\theta}_{\rm u} \approx \theta_{\rm u}$, then using \eqref{eq:hnoisegen} and \eqref{model_1}, we can observe that 
%\begin{align} \label{bdef1}
% \mathbf{b}_{i} &= g \, N_{\rm u}\mathbf{a}_{\rm B}(\theta_{\rm b})\mathbf{a}_{\rm R}^{H}(\phi_{\rm b},\psi_{\rm b}) \boldsymbol{\Omega}_{i} \mathbf{a}_{\rm R}(\phi_{\rm u},\psi_{\rm u}) + \boldsymbol{\nu}_{i}, \nonumber \\
% &= g{\bf v}_i + \boldsymbol{\nu}_{i},  \> i=1,2,\ldots,L,       
%\end{align}
%where $\boldsymbol{\nu}_{i} = \mathbf{W}_{i}\mathbf{a}_{{\rm u}}(\hat{\theta}_{\rm u})$ is the noise term.
%
By defining the length-$LN_{\rm b}$ vectors
$\mathbf{b} = [{\bf b}_1^T,\ldots, {\bf b}_L^T]^T$, ${\mathbf{v}} = [{\bf v}_1^T,\ldots, {\bf v}_L^T]^T$, and ${\bf z} = [{\bf z}_{1}^{T},\ldots,{\bf z}_{L}^{T}]^{T}$, we can estimate the complex path gain in the least squares sense as
\begin{equation} \label{gamma_est}
    \hat{g} = \frac{{\mathbf{v}}^{H}\mathbf{b}}{\vert \vert {\mathbf{v}} \vert \vert ^{2}},
\end{equation}
where we use the angle estimates $({\phi}_{\rm b},{\psi}_{\rm b}) = (\hat{\phi}_{\rm b},\hat{\psi}_{\rm b})$ to compute ${\bf v}$ and ${\theta}_{\rm u} = \hat{\theta}_{\rm u}$ to compute ${\bf b}$.

\section{Numerical experiments}  \label{sec:numerical_simulations}

\begin{figure}[t] 
\centering
  \includegraphics[width=0.5\columnwidth]{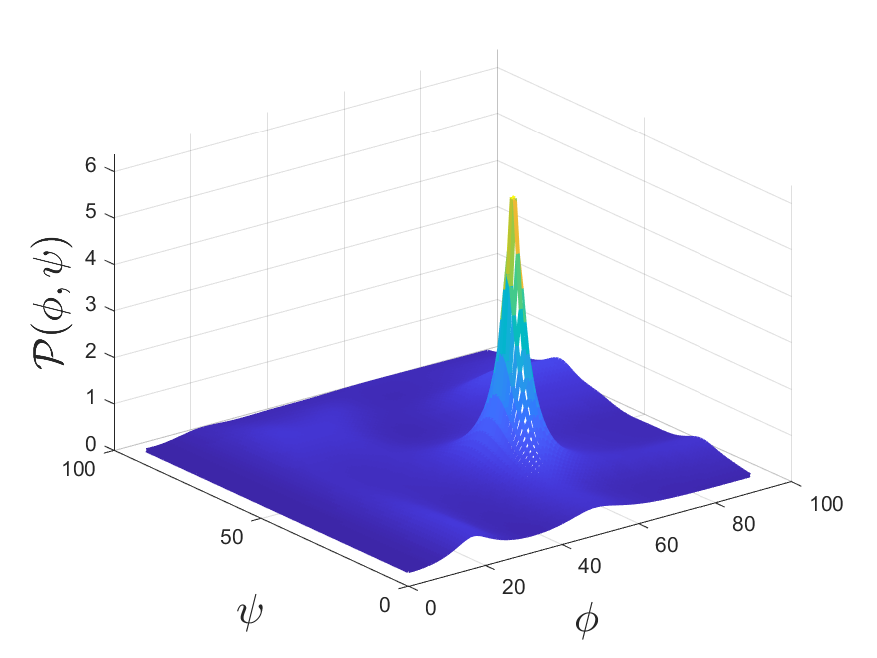}
 \vspace*{-2mm}
   \caption{Pseudo  spectrum~\eqref{eq:phi_psi_est} with  $L= 5$, $N_{\rm r} = 16 \times 16$, ${\rm SNR} = -10~{\rm dB}$.} 
   \label{fig1:phi_psi_spec} 
\vspace*{-6mm}
\end{figure}
\begin{figure*}[t]
%\vspace*{-1mm}
\begin{subfigure}[c]{0.31\columnwidth}\centering
   \includegraphics[width=\columnwidth]{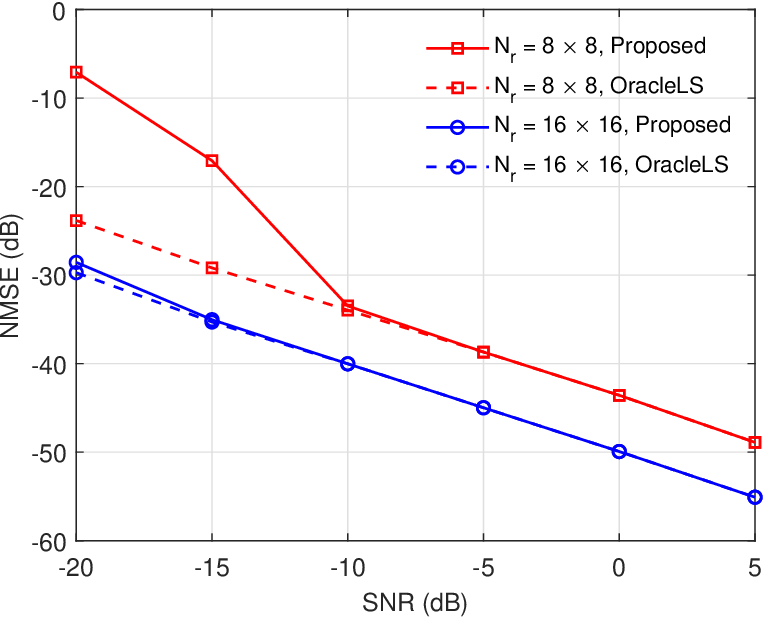}
   \caption{}
   \label{fig2:MSEvssnr}
\end{subfigure}
~
\begin{subfigure}[c]{0.31\columnwidth} \centering
   \includegraphics[width=\columnwidth]{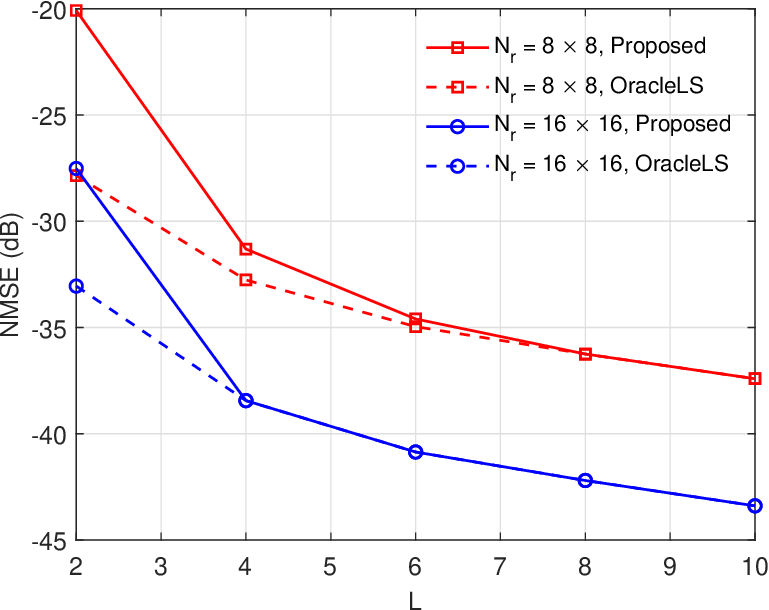}
   \caption{}
   \label{fig3:MSEvsL} 
\end{subfigure}
~
\begin{subfigure}[c]{0.31\columnwidth} \centering
   \includegraphics[width=\columnwidth]{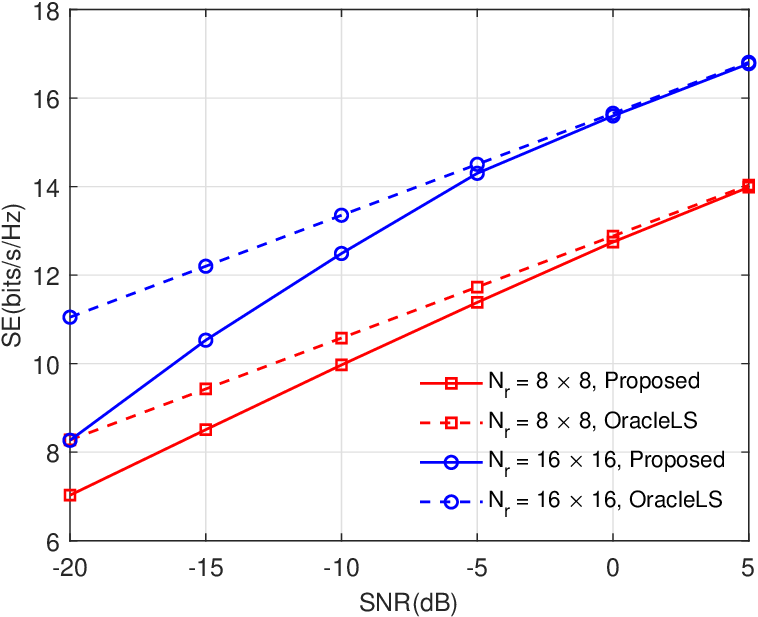}
   \caption{}
   \label{fig4:SEvsSNR}
\end{subfigure}
\vspace*{-2mm}
\caption{ (a) NMSE for different SNR with $L= 5$.
(b) NMSE for different $L$ at ${\rm SNR} = -10~{\rm dB}$. (c) Average SE for different SNR  for $L= 2$.}
\vspace*{-6mm}
\end{figure*}

In this section, we present simulation results to demonstrate
the effectiveness of the proposed channel estimation technique by computing the normalized mean square error (NMSE) and average spectral efficiency (SE). 
 For comparison, we use a scheme that assumes a perfect knowledge of the angles at the UE, RIS, and BS, thereby reducing the problem of channel estimation to that of estimating the overall complex path gain $g$. This method is referred to as the \textit{oracle least squares}~(\texttt{OracleLS}) estimator, which also forms the benchmark for compressed sensing based parametric channel estimation schemes~\cite{wang2020compressed,he2020channelirw,he2020channelanm}.
We define the NMSE  as
\begin{equation}
   {\rm NMSE} = \frac{\mathbb{E}\left[\| \mathbf{H}_{L} - \hat{\mathbf{H}}_{L} \|_{F}^{2}\right]}{\mathbb{E}\left[\| \mathbf{H}_{L} \|_{F}^{2} \right]},
\end{equation}
 where $\hat{\mathbf{H}}_{L} = \hat{g} \mathbf{a}_{\rm B}(\theta_{\rm b})\mathbf{a}_{\rm R}^{H}(\phi_{\rm b},\psi_{\rm b}) \boldsymbol{\Omega}_{L} \mathbf{a}_{\rm R}(\hat{\phi}_{\rm u},\hat{\psi}_{\rm u})\mathbf{a}_{\rm U}^{H}(\hat{\theta}_{\rm u})$.

%We note that the Cramer-Rao bound (CRB) for the channel estimation is always lower bounded by the CRB for the path gain estimation. Since \texttt{OracleLS} is the optimal estimator for $g$, MSE obtained using \texttt{OracleLS} can also be considered as the lower bound for CRB of channel estimation. 

Average SE indicates the performance of a RIS-assisted MIMO communication system, where the precoders at the UE and the phase shift matrix at the RIS are selected based on the estimated angles $\hat{\theta}_{\rm u}$, $\hat{\phi}_{\rm u}$, and $\hat{\psi}_{\rm u}$ to obtain the maximum transmit diversity and best possible reflection, respectively. The optimum choice of the phase shift matrix $\boldsymbol{\Omega} = {\rm diag}(\boldsymbol{\omega})$ to obtain the best possible reflection to focus the signal from the transmitter to the receiver is given  by
$ {\rm angle}([{\boldsymbol{\omega}}]_{i}) = {\rm angle}([\mathbf{a}_{\rm R}({\phi}_{\rm u},{\psi}_{\rm u})]_{i}) - {\rm angle}([\mathbf{a}_{\rm R}(\phi_{\rm b},\psi_{\rm b})]_{i})$~\cite{he2020channelirw}.  We now define the  average ${\rm SE} = \mathbb{E}\left[  \log(1 + \vert \mathbf{w}^{H} \mathbf{H}(\hat{\boldsymbol{\Omega}})\mathbf{f} \vert ^2 P) \right]$,
 %\begin{equation} \label{avg_se}
  %   {\rm SE} = \mathbb{E}\left[  \log(1 + \vert \mathbf{w}^{H} \mathbf{H}(\hat{\boldsymbol{\Omega}})\mathbf{f} \vert ^2 P) \right],
%\end{equation}
where  $\mathbf{w} = \frac{1}{\sqrt{N_{\rm b}}}\mathbf{a}_{\rm B}(\theta_{\rm b})$ is the combiner at the BS, $\mathbf{f} = \frac{1}{\sqrt{N_{\rm u}}}\mathbf{a}_{\rm U}(\hat{\theta}_{\rm u})$ is the optimum precoder at the UE, $\mathbf{H}(\hat{\boldsymbol{\Omega}}) = g \mathbf{a}_{\rm B}(\theta_{\rm b})\mathbf{a}_{\rm R}^{H}(\phi_{\rm b},\psi_{\rm b}) \hat{\boldsymbol{\Omega}} \mathbf{a}_{\rm R}(\phi_{\rm u},\psi_{\rm u})\mathbf{a}_{\rm U}^{H}(\theta_{\rm u})$, and $\hat{\boldsymbol{\Omega}} = {\rm diag}(\hat{\boldsymbol{\omega}})$ is the RIS phase matrix computed from the estimated angles.

We consider $N_{\rm u} = 8$ and $N_{\rm b} = 12$ throughout the simulations and the results are obtained by averaging over 1000 independent realizations of the receiver noise and unit-modulus complex path gain. An inter-element spacing of ${\lambda}/{2}$ is considered for the ULAs at the BS and UE. We have selected a sub-wavelength~\cite{najafi2020physicsbased} inter-element spacing of ${\lambda}/{5}$ for the UPA at the RIS. The phase shifts are selected randomly by uniformly sampling the unit circle. We considered a setup  with $\theta_{\rm b} = 40^{0}$, $\theta_{\rm u} =40 ^{0}$, $\phi_{\rm b} =50 ^{0}$, $\psi_{\rm b} = 65^{0}$, $\phi_{\rm u} = 50^{0}$, and $\psi_{\rm u} =30 ^{0}$. The angles $\theta_{\rm b}$, $\phi_{\rm b}$ and $\psi_{\rm b}$ are considered to be known from the knowledge of the positions of the RIS and BS. The remaining angles as well as the complex path gain are estimated.

%, thus leading to a sharper pseudo-spectrum

In Fig. \ref{fig1:phi_psi_spec}, we illustrate the pseudo spectrum $\mathcal{P}(\phi,\psi)$ in~\eqref{eq:phi_psi_est} for an SNR of $-10~{\rm dB}$ and for $L=5$. We can see that with multiple soundings, the pseudo spectrum results in a reasonable estimate of $(\phi_{u},\psi_{u})$ with a peak at the true location.  In Fig.~\ref{fig2:MSEvssnr}, we show NMSE for different SNRs and for different number of elements in the RIS, where we have used $L=5$. 
We can see that NMSE reduces with an increase in the number of RIS elements since the direction estimates improve with an increase in the array aperture.
%We can see that the NMSE reduces with an increase in the number of RIS elements with a larger aperture, which results in better direction estimates. 
In Fig. \ref{fig3:MSEvsL}, we show NMSE for different number of channel soundings, $L$, where we fix the SNR to $-10~\rm{dB}$. We can see that the channel estimation performance improves as $L$ increases. More importantly, for the considered setup, we can see that the algorithm performs similar to \texttt{OrcaleLS} for SNR above $-15~{\rm dB}$ and for $L>4$.

In Fig. \ref{fig4:SEvsSNR}, we show the average SE of \texttt{OrcaleLS} for different SNRs and different values of $N_{\rm r}$. Since the design of an optimal precoder, combiner, and phase shift matrix depend only on the angles and not on the path gains, the SE of the \texttt{OracleLS} estimator is the same as the maximum spectral efficiency  that can be achieved using a perfect channel state information. The proposed scheme achieves average SE as that of \texttt{OrcaleLS} for low SNRs (around $0~ {\rm dB}$) even when $L=2$. This means that the error in estimating the angles $\{\theta_{\rm   u}, \phi_{\rm   u}, \psi_{\rm   u}\}$  is very less. 

We can also observe that the performance of the proposed channel estimation scheme is improving with an increase in the number of elements (thus the aperture) of the RIS. We observe a 6 dB  and 3 dB gain in the NMSE and SE, respectively, when we quadruple the number of RIS elements from $N_{\rm r} = 8 \times 8$ to $N_{\rm r} = 16 \times 16$. This improvement in the  NMSE and SE is due to the improved angle estimates and higher beamforming gain with the larger RIS array.

%\vspace{-5mm}
\section{Conclusions} \label{sec:conclusions}

We have presented a pilot-based uplink channel estimation algorithm for RIS-assisted mmWave MIMO systems. We have assumed that the RIS-BS channel is known up to a complex path gain and considered an LoS angular channel model for the UE-RIS link. To resolve the ambiguity and uniquely estimate the complex path gain and the angles at the RIS, we have proposed a multiple channel sounding technique in which we observe the channel through different RIS phase shifts. We have presented an algorithm to estimate the RIS-UE channel parameters at the BS. Through numerical simulations, we have demonstrated that the proposed algorithm performs on par with a method that perfectly knows the locations of the BS, UE, and RIS.

%\newpage
\bibliographystyle{IEEEtran}
\bibliography{IEEEabrv,bibliography}

% Generated by IEEEtran.bst, version: 1.12 (2007/01/11)
\begin{thebibliography}{10}
\providecommand{\url}[1]{#1}
\csname url@samestyle\endcsname
\providecommand{\newblock}{\relax}
\providecommand{\bibinfo}[2]{#2}
\providecommand{\BIBentrySTDinterwordspacing}{\spaceskip=0pt\relax}
\providecommand{\BIBentryALTinterwordstretchfactor}{4}
\providecommand{\BIBentryALTinterwordspacing}{\spaceskip=\fontdimen2\font plus
\BIBentryALTinterwordstretchfactor\fontdimen3\font minus
  \fontdimen4\font\relax}
\providecommand{\BIBforeignlanguage}[2]{{%
\expandafter\ifx\csname l@#1\endcsname\relax
\typeout{** WARNING: IEEEtran.bst: No hyphenation pattern has been}%
\typeout{** loaded for the language `#1'. Using the pattern for}%
\typeout{** the default language instead.}%
\else
\language=\csname l@#1\endcsname
\fi
#2}}
\providecommand{\BIBdecl}{\relax}
\BIBdecl

\bibitem{rappaport2013millimeter}
T.~S. {Rappaport}, S.~{Sun}, R.~{Mayzus}, H.~{Zhao}, Y.~{Azar}, K.~{Wang},
  G.~N. {Wong}, J.~K. {Schulz}, M.~{Samimi}, and F.~{Gutierrez}, ``Millimeter
  wave mobile communications for 5{G} cellular: It will work!'' \emph{IEEE
  Access}, vol.~1, pp. 335--349, May 2013.

\bibitem{ghosh2014millimeter}
A.~{Ghosh}, T.~A. {Thomas}, M.~C. {Cudak}, R.~{Ratasuk}, P.~{Moorut}, F.~W.
  {Vook}, T.~S. {Rappaport}, G.~R. {MacCartney}, S.~{Sun}, and S.~{Nie},
  ``Millimeter-wave enhanced local area systems: A high-data-rate approach for
  future wireless networks,'' \emph{{IEEE} J. Sel. Areas Commun.}, vol.~32,
  no.~6, pp. 1152--1163, June 2014.

\bibitem{wan2020broadband}
Z.~Wan, Z.~Gao, and M.-S. Alouini, ``Broadband channel estimation for
  intelligent reflecting surface aided mm{W}ave massive {MIMO} systems,''
  \emph{arXiv preprint arXiv:2002.01629}, Feb. 2020.

\bibitem{mo2014channel}
J.~{Mo}, P.~{Schniter}, N.~G. {Prelcic}, and R.~W. {Heath}, ``Channel
  estimation in millimeter wave {MIMO} systems with one-bit quantization,'' in
  \emph{Proc. of the Asilomar Conference on Signals, Systems and Computers},
  Pacific Grove, USA, Nov. 2014.

\bibitem{basar2019wireless}
E.~{Basar}, M.~{Di Renzo}, J.~{De Rosny}, M.~{Debbah}, M.~{Alouini}, and
  R.~{Zhang}, ``Wireless communications through reconfigurable intelligent
  surfaces,'' \emph{IEEE Access}, vol.~7, pp. 116\,753--116\,773, Aug. 2019.

\bibitem{ozdogan2019intelligent}
{\"O}.~{\"O}zdogan, E.~Bj{\"o}rnson, and E.~G. Larsson, ``Intelligent
  reflecting surfaces: Physics, propagation, and pathloss modeling,''
  \emph{IEEE Wireless Commun. Lett.}, vol.~9, no.~5, pp. 581--585, May 2019.

\bibitem{ozdogan2020using}
------, ``Using intelligent reflecting surfaces for rank improvement in {MIMO}
  communications,'' in \emph{Proc. of the IEEE International Conference on
  Acoustics, Speech, and Signal Processing (ICASSP)}, Barcelona, Spain, May
  2020.

\bibitem{bjornson2020intelligent}
E.~Bj{\"o}rnson, {\"O}.~{\"O}zdogan, and E.~G. Larsson, ``Intelligent
  reflecting surface versus decode-and-forward: How large surfaces are needed
  to beat relaying?'' \emph{IEEE Wireless Commun. Lett.}, vol.~9, no.~2, pp.
  244--248, Feb. 2020.

\bibitem{najafi2020physicsbased}
M.~Najafi, V.~Jamali, R.~Schober, and H.~V. Poor, ``Physics-based modeling and
  scalable optimization of large intelligent reflecting surfaces,'' \emph{arXiv
  preprint arXiv:2004.12957}, Apr. 2020.

\bibitem{arun2019rfocus}
V.~Arun and H.~Balakrishnan, ``Rfocus: Practical beamforming for small
  devices,'' \emph{arXiv preprint arXiv:1905.05130}, May 2019.

\bibitem{taha2019enabling}
A.~Taha, M.~Alrabeiah, and A.~Alkhateeb, ``Enabling large intelligent surfaces
  with compressive sensing and deep learning,'' \emph{arXiv preprint
  arXiv:1904.10136}, Apr. 2019.

\bibitem{xiao2021low_complexity}
X.~{Chen}, J.~{Shi}, Z.~{Yang}, and L.~{Wu}, ``Low-complexity channel
  estimation for intelligent reflecting surface-enhanced massive {MIMO},''
  \emph{IEEE Wireless Commun. Lett.}, Jan. 2021.

\bibitem{wang2020compressed}
P.~Wang, J.~Fang, H.~Duan, and H.~Li, ``Compressed channel estimation for
  intelligent reflecting surface-assisted millimeter wave systems,''
  \emph{{IEEE} Signal Process. Lett.}, vol.~27, pp. 905--909, May 2020.

\bibitem{he2020channelirw}
J.~He, M.~Leinonen, H.~Wymeersch, and M.~Juntti, ``Channel estimation for
  {RIS}-aided mm{W}ave {MIMO} channels,'' \emph{arXiv preprint
  arXiv:2002.06453}, Feb. 2020.

\bibitem{he2020channelanm}
J.~He, H.~Wymeersch, and M.~Juntti, ``Channel estimation for {RIS}-aided
  mm{W}ave {MIMO} systems via atomic norm minimization,'' \emph{arXiv preprint
  arXiv:2007.08158}, July 2020.

\bibitem{he2020cascaded}
Z.-Q. He and X.~Yuan, ``Cascaded channel estimation for large intelligent
  metasurface assisted massive {MIMO},'' \emph{IEEE Wireless Commun. Lett.},
  vol.~9, no.~2, pp. 210--214, Feb. 2020.

\bibitem{vantrees2002optimum}
H.~L. Van~Trees, \emph{Optimum Array Processing: Part IV of Detection,
  Estimation, and Modulation Theory}.\hskip 1em plus 0.5em minus 0.4em\relax
  USA: John Wiley \& Sons, Ltd, 2002.

\end{thebibliography}

\end{document}